\documentstyle[epsf,12pt]{article}
\newcommand{\al}{\alpha'}

\newcommand{\de}{\partial}
\newcommand{\be}{\begin{equation}}
\newcommand{\ba}{\begin{eqnarray}}
\newcommand{\ea}{\end{eqnarray}}
\newcommand{\ee}{\end{equation}}

\newcommand{\f}{\frac}
\newcommand{\s}{\sqrt}

\newcommand{\ti}{\tilde}
\newcommand{\ap}{\alpha}

\newcommand{\mb}{\mathbf}
\newcommand{\ddd}{\cdot\cdot\cdot}

\newcommand{\no}{\nonumber \\}
\newcommand{\la}{\langle}
\newcommand{\lb}{\rangle}

\begin{document}
\baselineskip 6mm
\begin{titlepage}
\thispagestyle{empty}
%\begin{flushright}
%hep-th/0008023 \\
%UT-903 \\
%August, 2000 \\
%\end{flushright}

~\vspace{7ex}

\begin{center}
\noindent{\LARGE \textbf{Tachyon Condensation on Fuzzy Sphere}}\\
\smallskip
\noindent{\LARGE \textbf{and Noncommutative Solitons}}\\
\vspace{11ex}
\noindent{ Yasuaki Hikida,\footnote{
                 E-mail: hikida@hep-th.phys.s.u-tokyo.ac.jp}
           Masatoshi Nozaki\footnote{
                 E-mail: nozaki@hep-th.phys.s.u-tokyo.ac.jp} and
           Tadashi Takayanagi\footnote{
                 E-mail: takayana@hep-th.phys.s.u-tokyo.ac.jp} }\\
\bigskip
{\it Department of Physics, Faculty of Science \\ University of Tokyo \\
\medskip
Tokyo 113-0033, Japan}
\vspace{9ex}
\end{center}
\begin{abstract}
We study a brane-antibrane system and a non-BPS D-brane in $SU(2)$ WZW model.
We first discuss the tachyon condensation using the vertex operator formalism
and find the generation of codimension two D-branes after the condensation.
Our result is consistent with the recent interpretation that a D2-brane
is a bound state of D0-branes.
Then we investigate the world-volume effective theory on a non-BPS D-brane.
It becomes a field theory on the ``fuzzy sphere'' when the level is sent to infinity.
The most interesting feature is that there exist
the noncommutative tachyonic solitons and  we can identify them with D0-branes.
We also discuss the brane-antibrane system from the world-volume point of view
and comment on the relation to the noncommutative version of the index
 theorem.

~\\
PACS: 11.25.-w; 11.25.Hf\\
Keywords; Noncommutative Geometry, Tachyon Condensation\\
hep-th/0008023\\
UT-903
\end{abstract}
\end{titlepage}

\newpage

%\tableofcontents
\section{Introduction}
\setcounter{equation}{0}

The study of tachyon condensation \cite{ba,sen13} in open string theory
has been important to know the dynamics of non-BPS states in
superstring theory. Up to now active investigations have been made in
this area \cite{sen18}. In particular Sen and Witten pointed out 
 that a tachyon configuration on a non-supersymmetric brane system 
 generates a lower dimensional brane system if the configuration has a
non-trivial topology or K-theory charge \cite{sen14,witten1,olsen}. 
One way to verify this conjecture is to use the
description of the boundary conformal field theory
\cite{sen14,sen16,sen19,sen26,sen277} and their boundary state
descriptions were given in \cite{frau,NTU}. 

Non-BPS solitonic objects in
superstring theory are classified into two types. One is a
brane-antibrane system \cite{ba,sen13,sen14,sen16} and the other is a
non-BPS D-brane \cite{bergman2,sen14,sen16,sen22,gab}. Both systems
include open strings which have the opposite GSO-projection and the
tachyonic instability occurs in both cases. 

So far most of the examples in this subject have been
studied only in flat space or various orbifolds. As a next
step one would like to discuss the dynamics of non-BPS states which
consist of curved D-branes in more general backgrounds such as group
manifolds. Here we will consider D-branes in $SU(2)$ WZW
model. Such a background indeed appears in the near horizon
limit of NS5-branes \cite{chs}. A D-brane in $SU(2)$ WZW model is
described by a boundary state which satisfies the Cardy's condition
\cite{cardy}. It describes a D2-brane wrapping on a conjugacy class of
$SU(2)$ \cite{alek1}. Moreover the low energy effective theory on the brane
turns out to be a noncommutative theory because of the background
$H$-flux \cite{alek2,alek3}. In particular the algebraic structure of
this theory is identified with that of a fuzzy sphere \cite{fuzzy,fuzzy2}
in the infinite level limit. 

 The purpose of this paper is to study the tachyon condensation on a
 non-BPS D-brane or a pair of brane-antibrane in $SU(2)$ WZW model from
 two different points of view. The first approach is to investigate the vertex
 operators which correspond to various tachyonic modes. In the
 discussion below we will
 determine what is generated after the tachyon condensation when we apply
 the Sen's conjecture. Interestingly the results are consistent with the
 bound state interpretation argued in \cite{douglas,paw}. The second is
 to study the low energy effective action of the tachyon field. 
 In this approach it is crucial to take the infinite level limit, 
 which corresponds to the infinite $B$-field limit. 
 The notable point is that the effective action is dramatically simplified, 
 which reflects the fact that the world-volume of a D-brane
 can be regarded as a fuzzy sphere. Consequently
  this leads to the interesting
 observation that an excitation of the tachyon field on the non-BPS
 D-brane is very much like the noncommutative soliton \cite{gop}.
 This can be seen as a ``noncommutative tachyon"
 \cite{harvey,das,witten2} on the fuzzy sphere. Indeed we can see that
 this soliton approaches the known noncommutative soliton in a flat space
 in the large world-volume limit. 

The paper is organized as follows. In section 2 we first review 
the boundary state description of D-branes in $SU(2)$ WZW model. After
that we investigate the tachyon vertex operators and discuss the
generation of  
codimension two D-branes on the world-volume of a pair of
brane-antibrane. In section 3 we discuss excitations of the tachyon
field in the effective theory on a non-BPS D-brane and show that these
can be identified with ``noncommutative solitons" if we take the limit of
level infinity. We also consider the brane-antibrane
system and mention the relation between the tachyon condensation and the
index theorem of a noncommutative algebra.

\section{Open strings in $SU(2)$ WZW model and tachyon condensation}
\setcounter{equation}{0}

In this section first we review the description of open strings between
D-branes in $SU(2)$ WZW model and then discuss tachyonic modes on
brane-antibrane systems and non-BPS D-branes. 
%
%In this paper we will be
%mainly interested in the critical superstring theory and in this case 
%
$SU(2)$ WZW model with level $k$ appears naturally in the near-horizon
geometry of 
$N$ parallel NS 
$5$-branes \cite{chs} when one is interested in the critical
superstring theory; the relation between the level $k$ and $N$ is
given by $N=k+2$. The total geometry is  
\be
{\mb{R}}^{5,1}\times {\mb{R}}_{\phi}\times {\mb{S}}^3,
\ee
where ${\mb R}_{\phi}$ is the radial direction.

This radial direction is described by the
linear dilaton theory. The boundary condition along this
direction is Neumann or Dirichlet \cite{li}. Under the Neumann 
boundary condition, the worldsheet fermions are the same as in the flat
space. Therefore D-branes preserve some supersymmetries. 
On the other hand, the Dirichlet boundary
condition seems to break supersymmetry 
%and be unstable 
and the explicit Dirichlet boundary state 
in supersymmetric theories
has not been known yet.

Throughout this paper we will omit the ${\mb{R}}^{5,1}\times
{\mb{R}}_{\phi}$ geometry and study explicitly only the ${\mb{S}^3}$ geometry. 
We also assume that the coupling constant in the string theory is so small
that we can trust the tree level analysis. 

\subsection{D-branes in $SU(2)$ WZW model}

In rational conformal field theories which have the diagonal modular
invariance, one can construct the boundary states of D-branes by the
Cardy's prescription \cite{cardy} in general. The result is 
\ba
|p\lb=\sum_{a=1}^{N-1}\f{S_{p a}}{\s{S_{1 a}}}|a\lb\lb\ \ \ \ \ (1\leq p \leq N-1),
\ea
where $|a\lb\lb$ is called the Ishibashi state \cite{ishi} which
corresponds to the highest weight state of spin $\f{a-1}{2}$; the
modular transformation matrix is denoted by
\ba
S_{p a}=\s{\f{2}{N}}\sin(\f{pa\pi}{N}).
\ea

The geometrical interpretation of these $D$-branes was first given in
\cite{alek1} and further discussed\footnote{For the earlier discussions
of this subject, see \cite{sag,kato}.} in
\cite{alek2,alek3,douglas,paw,geo,giveon}. $SU(2)$ WZW model can
be interpreted as a sigma model of which target space is $SU(2)\simeq \mb{S}^3$
with $N=k+2$ units of $H$-flux.
The shift of level by two is due to the contribution of fermionic sector
 and the radius of $\mb{S}^3$ is $\s{N\al}$. Let us define the
coordinates of ${\mb{S}}^3$ as
\ba
(\cos\psi,\sin\psi\cos\phi\sin\theta,\sin\psi\sin\phi\sin\theta,\sin\psi\cos\theta)\in
{\mb{R}}^4.
\ea
Then a D-brane $|p\lb$ can be identified with
the one wrapping on the $p$-th conjugacy class of $SU(2)$, which corresponds to
%In other words the world-volume of the D-brane is 
$\mb{S}^2\subset
\mb{S}^3$ with $\psi=\f{p\pi}{N}$ \cite{alek1}. 
A D-brane corresponding to $p=1$ or $p=N-1$ is equivalent to a pure
D0-brane and the other branes can be regarded as D2-branes. Even
though these D2-branes wrap on the topologically trivial 2-cycles, they
are stabilized by the combined effect of the $H$-flux and the gauge flux 
\ba
F=-\f{p}{2}\sin\theta d\theta d\phi \label{flux}
\ea
on the world-volume \cite{douglas}. This shows that such a D-brane can
be interpreted as a bound state of $p$ D0-branes \cite{douglas} and it
was pointed out  in \cite{alek3} that a stuck of D0-branes can condense
into a single D2-brane. 

The open string spectrum between a pair of D-branes corresponding to $|p\lb$ and $|q\lb$ is represented as the following cylinder amplitude
\ba
Z_{p q}(q)&=&\la p|\tilde{q}^{L_0-\f{c}{24}}|q\lb
~=~\sum_{j}n^j_{p q}\chi_j(q),\\
\chi_j(q)&=&\mbox{Tr}_{j}(q^{L_0-\f{c}{24}}),\nonumber
\ea
where $\ti{q}=e^{-2\f{\pi}{t}}$ and $q=e^{-2\pi t}$ are closed string
and open string moduli of the cylinder, respectively. The character
$\chi_j(q)$ means the partition function of the open string which belongs to spin $j$ sector. 
A notable point is the appearance of the fusion coefficient $n^j_{p q}\in
\mb{Z}$ as discussed in \cite{cardy} . In the case of $SU(2)$ WZW model,  $n^j_{p q}$ is given as follows
\ba 
n^j_{p q}&=&1\ \  \mbox{if}\ \ \f{|p-q|}{2}\leq j \leq \mbox{min}\{\f{p+q}{2}-1,\ N-1-\f{p+q}{2}\}, \nonumber \\
n^j_{p q}&=&0 \ \  \mbox{elsewhere}\ .
\ea

Then the lightest modes or equally the zero modes of the open strings can be represented as the following vertex operators
\ba
V_{j,m}\ :\ -j\leq m \leq j\ ,\ \ \f{|p-q|}{2}\leq j \leq \mbox{min}\{\f{p+q}{2}-1,\ N-1-\f{p+q}{2}\}.
\ea
Their conformal dimensions are given by
\ba
\Delta_{j,m}=\f{j(j+1)}{N}.
\ea

Now we turn to the ten dimensional string theory including the
non-compact directions and assume the
Neumann boundary condition for ${\mb{R}}_{\phi}$ direction.  Then we obtain
the mass spectrum for the above vertex operators as
\ba
\al m^2=-\f{1}{2}+\f{j(j+1)}{N}+\f{Q^2}{8}, \label{masf}
\ea
where $Q=\s{\f{2}{N}}$ is the background charge for the linear dilaton
sector. Here we used normalizable states, of which
spectrum has the lower bound $\frac{Q^2}{8}$ and its description by using $SL(2,{\mb R})/U(1)$ model was discussed in \cite{giveon}. 
To see the lowest mass states, we assumed that 
there is no momentum along any coordinates except $\mb{S}^3$.     
From this we can see that the term $\frac{Q^2}{8}=\f{1}{4N}$  does not change
the qualitative behavior of tachyon fields. In particular if we take
$N\to\infty$ limit, then we can neglect this. 
Though the explicit form of the Dirichlet boundary state has not been known
 in the supersymmetric linear dilaton theory, we assume that the
similar behavior will occur. 

\subsection{Tachyon condensation in $SU(2)$ WZW model}

Let us turn to the tachyon condensation on a pair of brane-antibrane or on
a non-BPS D-brane in $SU(2)$ WZW model. As explained in the previous
subsection, only D2-branes ($|2\lb,|3\lb,\ddd,|N-2\lb$) and D0-branes
($|1\lb,|N-1\lb$) are allowed in this model whether the system is
BPS or not. Here we assume $N$ is finite and therefore the conformal
dimensions of the vertex operators with different spins $j$ are not the same.

First we consider a brane-antibrane system where both branes correspond
to the boundary state $|p\lb$\footnote{Notice that an
antibrane have the same bosonic part of the boundary state as a brane.
However if one
take the fermionic part into consideration, then each has a different
sign of the RR-sector.}. In order to distinguish them, we denote the
brane by $|p\lb$ and the antibrane by $|\bar{p}\lb$. We can assume
$1\leq p\leq \f{N}{2}$ without loss of generality. Then the open strings
between D-brane and anti D-brane have the opposite GSO-projection and
the ``tachyonic mode'' appears once for each vertex operator $V_{j,m}\ \
(j=0,1,\ddd,p-1)$. If we assume $p< \s{\f{N}{2}}$, then all of these
modes are really tachyonic ($m^2<0$). Even if $p > \s{\f{N}{2}}$,
these operators do not include any oscillator excitations and 
can be considered to belong to the sector of tachyon field. Then let us
give an interpretation of these modes from the viewpoint of the
world-volume theory. The world-volume of the brane-antibrane system is  
$\mb{S}^2$ and its radius is $\s{N\al}\sin\f{\pi p}{N}$. The low energy
effective theory consists of the (complex) tachyon field and the
massless fields (a gauge field and a scalar field). The crucial
observation is that the tachyonic mode $V_{j,m}$ in string theory
corresponds to the tachyon field on the world-volume which is
proportional to the spherical harmonics $Y_{j,m}(\theta,\phi)$. This 
interpretation is very natural as implied in \cite{geo,alek3} and an
elementary explanation is as follows. If one restricts the
$SU(2)$ currents in the WZW model to zero modes $(\theta,\phi)$ on the
sphere and quantizes the modes, then one obtains the conventional angular
momentum operator in quantum mechanics. For example, the mode $V_{j,j}$
represents the tachyon field  
\be
T(\theta,\phi)=T_{0}\ e^{ij\phi}(\sin\theta)^j,
\ee
where $T_{0}$ is a constant. Note that this field has two nodes
$T(\theta,\phi)=0$ at $\theta=0,\pi$ and they correspond to the vortex
line configuration of winding number $j,-j$, respectively.

In general, $n$ codimension two D-branes are generated after the tachyon 
condensation of winding number $n$ vortex line
\cite{sen14,witten1,sen26,NTU}. Thus we can conclude that the
condensation of a tachyonic mode $V_{j,j}$ produces $j$ D0-branes at
$\theta=0$ and $j$ anti D0-branes at $\theta=\pi$. This result is
consistent with the allowed values $j=0,1,\ddd,p-1$, because the original
(anti) D2-brane $|p\lb$ ($|\bar{p}\lb$) can be thought as a bound state
of $p$ (anti) D0-branes as we mentioned above. For example,
 the condensation of
$j=p-1$ mode means the annihilation of only one D0 and anti D0-brane;
the constant mode $j=0$ corresponds to the annihilation of all
branes and the system will eventually go down to the vacuum. 
Consequently we can say that this 
gives another evidence of the bound state interpretation discussed in
\cite{douglas,alek3}. The similar argument is applicable if
$m=-j$. 

Next let us discuss the mode $V_{j,m}\ \ (|m|\neq j)$. Naively
one may think that because the corresponding tachyon field proportional
to $Y_{j,m}$ includes one dimensional nodes $Y_{j,m}=0$, its condensation
will produce codimension one D-branes according to the Sen's
conjecture. However it is easy to see that one can not construct boundary
states of such D-branes in our model. We argue that the solution to this
puzzle is the following. If we condense the tachyonic mode
$V_{j,m}$(or equally give the relevant perturbation in
the sense of \cite{rel}), then the OPEs of more than three insertions
at the worldsheet boundary include the mode $V_{m,m}$ because of
the conservation of the angular momentum. Since this mode is more
relevant than $V_{j,m}$, the condensation of $V_{j,m}$ results in the
previous case and the generations of codimension one D-branes do not
really occur. 

Then we turn to the case where a D-brane corresponds to $|p\lb$ and an
anti D-brane to $|\bar{q}\lb\ \ (p\neq q)$. Without losing generality,
we can assume $p >q$ and $\f{p+q-2}{2}\leq N-1-\f{p+q}{2}$. Then the open
string spectrum of zero modes is given by $V_{j,m}\
(j=\f{p-q}{2},\f{p-q+2}{2},\ddd,\f{p+q-2}{2})$. At first sight one may
be in trouble with the fact that if $p-q$ is odd, then one will get the
``double valued spherical harmonics" and fractional
D0-branes seem to be generated after the tachyon condensation.  
To resolve this one should note the crucial point that the open string
between these branes is affected by the gauge
flux $F^{(p)},\ F^{(q)}$ on each world-volume. In other words the
tachyonic field $T(\theta,\phi)$ in this case is not a function on
${\mb{S}}^2$ but a section of a line bundle which couples to the
difference of the gauge fields 
\ba
A^{(p)}-A^{(q)}=\f{p-q}{2}\cos\theta d\phi.
\ea
Therefore the tachyon field is affected by the gauge holonomy as follows
\ba
T(\theta,\phi)=T_0\cdot Y_{j,m}\cdot e^{i\int_{0}^{\phi}d\phi (A_{\phi}^{(p)}-A_{\phi}^{(q)})}=T_0\cdot Y_{j,m}e^{i\f{p-q}{2}cos{\theta}\cdot\phi},
\ea
where $T_0$ is a constant again.
Thus we get $(m+\f{p-q}{2})$ D0-branes at $\theta=0$ and
$(m-\f{p-q}{2})$ anti D0-branes at $\theta=\pi$ without any emergence of
the fractional D0-branes. For instance  if we consider\footnote{Here
again we have only to consider the cases $m=j,-j$ as in the previous
discussion.} the lowest mode $j=m=\f{p-q}{2}$, then we get $(p-q)$
D0-branes only at $\theta=0$. In the case of the highest mode
$j=m=\f{p+q}{2}-1$ the tachyon condensation generates $(p-1)$ D0-branes
at $\theta=0$ and $(q-1)$ anti D0-branes at $\theta=\pi$ and this
corresponds to one pair annihilation. 
These observations are again consistent with the interpretation that a D2-brane of type $|p\lb$ is a bound state of $p$ D0-branes. 

Finally let us turn to the tachyon condensation on a non-BPS D2-brane in superstring or a D2-brane in bosonic string. In these cases the tachyon field is a 
real scalar field and only the linear combinations such as
$V_{j,m}+V_{j,-m},\ i(V_{j,m}-V_{j,-m})$ are allowed. Therefore we
cannot gain the vortex line configurations and it is hard to tell what
is generated after the condensation. We will investigate this issue in
the next section from a different point of view. 

\section{Tachyon condensation on fuzzy sphere and noncommutative solitons}
\setcounter{equation}{0}

In this section we investigate the tachyon condensation from the
viewpoint of the world-volume effective theory. 
 Here we take the limit $N\to\infty$\footnote{Roughly speaking,
this corresponds to the large $B$-field limit which is taken in
\cite{harvey,das,witten2}, where D-branes in a flat space are considered. Quite
recently, the construction of the noncommutative soliton without taking
this limit is discussed in \cite{gop2,zhou}.} and in this limit the algebra
of vertex operators in $SU(2)$ WZW model is equivalent
to that of the fuzzy sphere \cite{fuzzy,fuzzy2} as shown in
\cite{alek2}. 
A fuzzy sphere is defined by identifying the following $SU(2)$ algebra
with its coordinates 
\ba
[X^i,X^j]=i\f{2}{\s{p^2-1}}\epsilon_{ijk}X^k,\ \ \  (X^1)^2+(X^2)^2+(X^3)^2=1,
\ea
where $X^i\ (i=1,2,3)$ are $p\times p$ matrices.
The integer $p$ labels the algebra of fuzzy sphere and
it can be identified with the label $p$ of D2-brane.

In the discussion below, we will study the
tachyon condensation on a non-BPS D-brane, which generates codimension
two non-BPS D-branes and then investigate the brane-antibrane
systems.

The effective action which describes the gauge theory on a (BPS)
D2-brane was constructed in \cite{alek3} and its interesting structure
was explicitly shown. Now we would like to consider the world-volume
theory on a non-BPS D2-brane corresponding to $|p\lb$. This includes
a real tachyon field $T$ which is described by a (quantum mechanical)
$p\times p$ hermitian matrix. In the large radius limit ($p\to\infty$) the
world-volume approaches a flat space and the dynamics is described by a
noncommutative field theory. Its action is given by 
\ba
S=\f{1}{G_s}\int dt(dx)^2\s{G}[G^{\mu\nu}\de_{\mu}T\de_{\nu}T+V(*T)], \\
V(*T)=G_{s}T_{D2}-\f{1}{2\al}T*T+\ddd \nonumber.
\ea
Here we used the open string metric $G^{\mu\nu}$, the effective open
string coupling $G_s$ and the star product
$A*B(x)=e^{\f{1}{2}\Theta^{ij}\de_i\de_j'}A(x)B(x')|_{x'=x}$ as in the
conventional noncommutative field theory description \cite{sw}. Also we
defined $T_{Dp}$ as a Dp-brane tension. 

When we take finite $p$ (i.e. the world-volume is a fuzzy sphere) we obtain
the ``regularized version" of the above action as follows 
\ba
S&=&\f{2\pi\Theta}{G_s}\int dt\s{G}\mbox{Tr}\left[f(T)\sum_{i=1,2,3}T[X^i,[X^i,T]]+V(T)\right], \label{tak} 
\ea
where $\Theta$ denotes the noncommutative parameter \cite{sw} and its explicit value will be discussed later. Here the previous star-product is replaced with the ordinary product of matrices and the derivatives are represented as 
commutators. Note also that we abbreviate the higher derivative terms because we are interested in the limit $N\to \infty$. 

In general it is difficult to know the explicit form of $f(T)$.
Nevertheless we can determine its value before the tachyon condensation in
such a way that it is consistent with the mass formula (\ref{masf}), namely 
\ba
f(T=0)=\f{p^2-1}{4N\al}.
\ea 

It is easy to see that in the limit $N\to\infty$ the kinetic term $\sim
\f{j(j+1)}{N\al}$ is negligible and only the potential terms are
relevant. Note that there is $U(p)$ symmetry $T\to gTg^{-1},\ \ g\in
U(p)$ in this limit, which corresponds to the diffeomorphism of the
fuzzy sphere. By using this symmetry we can diagonalize the matrix
$T$. 

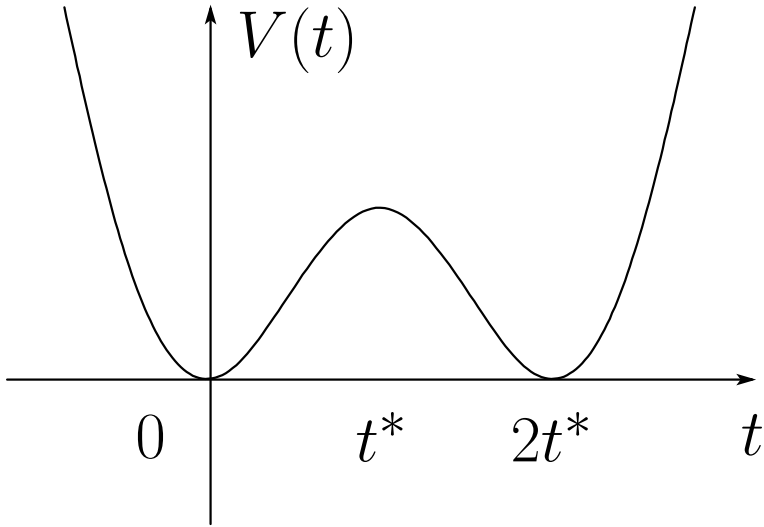
\begin{figure}[tbp]
\begin{center}
%WinTpicVersion2.15
\unitlength 0.1in
\begin{picture}(39.00,27.08)(4.00,-31.08)
% VECTOR 1 0 3 0
% 2 400 2750 4300 2750
% 
\special{pn 13}%
\special{pa 400 2350}%
\special{pa 4300 2350}%
\special{fp}%
\special{sh 1}%
\special{pa 4300 2350}%
\special{pa 4233 2330}%
\special{pa 4247 2350}%
\special{pa 4233 2370}%
\special{pa 4300 2350}%
\special{fp}%
% VECTOR 1 0 3 0
% 2 1465 3508 1465 808
% 
\special{pn 13}%
\special{pa 1465 3108}%
\special{pa 1465 408}%
\special{fp}%
\special{sh 1}%
\special{pa 1465 408}%
\special{pa 1445 475}%
\special{pa 1465 461}%
\special{pa 1485 475}%
\special{pa 1465 408}%
\special{fp}%
% SPLINE 1 0 3 0
% 5 700 800 1600 2675 2350 1850 3100 2675 4000 800
% 
\special{pn 13}%
\special{pa 700 400}%
\special{pa 711 452}%
\special{pa 723 504}%
\special{pa 734 555}%
\special{pa 745 607}%
\special{pa 757 658}%
\special{pa 768 710}%
\special{pa 780 761}%
\special{pa 791 812}%
\special{pa 803 862}%
\special{pa 814 913}%
\special{pa 826 963}%
\special{pa 837 1012}%
\special{pa 849 1061}%
\special{pa 860 1110}%
\special{pa 872 1159}%
\special{pa 884 1207}%
\special{pa 896 1254}%
\special{pa 908 1301}%
\special{pa 920 1347}%
\special{pa 932 1393}%
\special{pa 944 1438}%
\special{pa 956 1482}%
\special{pa 968 1525}%
\special{pa 981 1568}%
\special{pa 993 1610}%
\special{pa 1005 1651}%
\special{pa 1018 1692}%
\special{pa 1031 1731}%
\special{pa 1044 1770}%
\special{pa 1056 1807}%
\special{pa 1069 1844}%
\special{pa 1083 1880}%
\special{pa 1096 1914}%
\special{pa 1109 1948}%
\special{pa 1123 1980}%
\special{pa 1136 2011}%
\special{pa 1150 2041}%
\special{pa 1164 2070}%
\special{pa 1178 2098}%
\special{pa 1192 2124}%
\special{pa 1206 2149}%
\special{pa 1220 2173}%
\special{pa 1235 2195}%
\special{pa 1250 2216}%
\special{pa 1265 2236}%
\special{pa 1280 2254}%
\special{pa 1295 2271}%
\special{pa 1310 2286}%
\special{pa 1326 2299}%
\special{pa 1341 2311}%
\special{pa 1357 2321}%
\special{pa 1373 2330}%
\special{pa 1390 2337}%
\special{pa 1406 2342}%
\special{pa 1423 2345}%
\special{pa 1439 2347}%
\special{pa 1456 2346}%
\special{pa 1474 2344}%
\special{pa 1491 2340}%
\special{pa 1509 2334}%
\special{pa 1527 2327}%
\special{pa 1545 2317}%
\special{pa 1563 2305}%
\special{pa 1581 2291}%
\special{pa 1600 2275}%
\special{pa 1619 2257}%
\special{pa 1638 2237}%
\special{pa 1658 2215}%
\special{pa 1677 2191}%
\special{pa 1697 2166}%
\special{pa 1717 2140}%
\special{pa 1737 2112}%
\special{pa 1758 2083}%
\special{pa 1778 2054}%
\special{pa 1799 2023}%
\special{pa 1820 1992}%
\special{pa 1841 1961}%
\special{pa 1862 1929}%
\special{pa 1884 1897}%
\special{pa 1905 1865}%
\special{pa 1927 1833}%
\special{pa 1948 1801}%
\special{pa 1970 1770}%
\special{pa 1992 1739}%
\special{pa 2014 1709}%
\special{pa 2037 1680}%
\special{pa 2059 1652}%
\special{pa 2081 1626}%
\special{pa 2104 1600}%
\special{pa 2126 1576}%
\special{pa 2149 1554}%
\special{pa 2171 1533}%
\special{pa 2194 1514}%
\special{pa 2217 1498}%
\special{pa 2240 1483}%
\special{pa 2262 1471}%
\special{pa 2285 1462}%
\special{pa 2308 1455}%
\special{pa 2331 1451}%
\special{pa 2354 1450}%
\special{pa 2377 1452}%
\special{pa 2399 1457}%
\special{pa 2422 1465}%
\special{pa 2445 1475}%
\special{pa 2468 1488}%
\special{pa 2491 1503}%
\special{pa 2513 1520}%
\special{pa 2536 1540}%
\special{pa 2559 1561}%
\special{pa 2581 1584}%
\special{pa 2604 1608}%
\special{pa 2626 1634}%
\special{pa 2648 1662}%
\special{pa 2671 1690}%
\special{pa 2693 1719}%
\special{pa 2715 1749}%
\special{pa 2737 1780}%
\special{pa 2759 1812}%
\special{pa 2780 1843}%
\special{pa 2802 1875}%
\special{pa 2823 1907}%
\special{pa 2845 1939}%
\special{pa 2866 1971}%
\special{pa 2887 2002}%
\special{pa 2908 2033}%
\special{pa 2928 2063}%
\special{pa 2949 2093}%
\special{pa 2969 2121}%
\special{pa 2989 2149}%
\special{pa 3009 2175}%
\special{pa 3029 2199}%
\special{pa 3049 2222}%
\special{pa 3068 2243}%
\special{pa 3087 2263}%
\special{pa 3106 2280}%
\special{pa 3125 2296}%
\special{pa 3143 2309}%
\special{pa 3161 2320}%
\special{pa 3179 2329}%
\special{pa 3197 2337}%
\special{pa 3215 2342}%
\special{pa 3232 2345}%
\special{pa 3249 2347}%
\special{pa 3266 2346}%
\special{pa 3283 2344}%
\special{pa 3299 2340}%
\special{pa 3316 2335}%
\special{pa 3332 2327}%
\special{pa 3348 2318}%
\special{pa 3364 2307}%
\special{pa 3379 2295}%
\special{pa 3395 2281}%
\special{pa 3410 2265}%
\special{pa 3425 2248}%
\special{pa 3440 2230}%
\special{pa 3455 2210}%
\special{pa 3470 2188}%
\special{pa 3484 2166}%
\special{pa 3499 2141}%
\special{pa 3513 2116}%
\special{pa 3527 2089}%
\special{pa 3541 2061}%
\special{pa 3555 2032}%
\special{pa 3568 2001}%
\special{pa 3582 1970}%
\special{pa 3595 1937}%
\special{pa 3609 1903}%
\special{pa 3622 1868}%
\special{pa 3635 1832}%
\special{pa 3648 1795}%
\special{pa 3661 1757}%
\special{pa 3673 1718}%
\special{pa 3686 1679}%
\special{pa 3699 1638}%
\special{pa 3711 1597}%
\special{pa 3724 1554}%
\special{pa 3736 1511}%
\special{pa 3748 1467}%
\special{pa 3760 1423}%
\special{pa 3772 1378}%
\special{pa 3784 1332}%
\special{pa 3796 1286}%
\special{pa 3808 1239}%
\special{pa 3820 1191}%
\special{pa 3832 1143}%
\special{pa 3843 1094}%
\special{pa 3855 1045}%
\special{pa 3867 996}%
\special{pa 3878 946}%
\special{pa 3890 896}%
\special{pa 3901 846}%
\special{pa 3913 795}%
\special{pa 3924 744}%
\special{pa 3936 693}%
\special{pa 3947 642}%
\special{pa 3958 590}%
\special{pa 3970 538}%
\special{pa 3981 487}%
\special{pa 3992 435}%
\special{pa 4000 400}%
\special{fp}%
\special{pa 4000 400}%
\special{sp}%
% STR 2 0 3 0
% 3 1600 725 1600 800 1 0
% \Huge $V(t)$
\put(16.0000,-4.0000){\makebox(0,0)[lt]{\Huge $V(t)$}}%
% STR 2 0 3 0
% 3 2358 2975 2358 3050 5 0
% \Huge $t^*$
\put(23.5800,-26.5000){\makebox(0,0){\Huge $t^*$}}%
% STR 2 0 3 0
% 3 3250 2975 3250 3050 5 0
% \Huge $2 t^*$
\put(32.5000,-26.5000){\makebox(0,0){\Huge $2 t^*$}}%
% STR 2 0 3 0
% 3 1150 2975 1150 3050 5 0
% \Huge $0$
\put(11.5000,-26.5000){\makebox(0,0){\Huge $0$}}%
% STR 2 0 3 0
% 3 4300 2940 4300 3040 5 0
% \Huge $t$
\put(43.0000,-26.4000){\makebox(0,0){\Huge $t$}}%
\end{picture}%
\caption{Tachyon potential on the non-BPS D-brane.}
\label{doublewell}
\end{center}
\end{figure}

Further we assume the double well potential of a non-BPS D-brane and
shift the value of $T$ as $T\to T+t^*$ for convenience. Here we define
the local maximum of the function $V(t),\ t\in {\mb R}$ as $t^*$ (see figure
\ref{doublewell}).   This does not
change the kinetic term and the potential term behaves as
$V(T)=G_{s}T_{D2}-\f{1}{2\al}(T-t^*)^2+\ddd$.
As we will see, we 
can describe the tachyon condensation without knowing the detailed form
of the potential.  

Then we can describe the lightest excitation as
$T=t^*\mbox{diag}(0,\ddd,0,1,0,\ddd,0)$. 
Note that this matrix is a projection in the algebra of fuzzy sphere.
We claim that the excitation is
regarded as a ``noncommutative
soliton" \cite{gop} in a finite dimensional space; their applications to
the tachyon condensation 
have been discussed in \cite{harvey,das,witten2} recently. The mass of
the original D2-brane is known to be $p$ times as heavy as that of a
D0-brane in the limit $N\to\infty$ from the analysis of the boundary
state \cite{douglas}. Note also that the original D2-brane corresponds to
$T=t^*\mbox{diag}(1,1,\ddd,1)$ in our model. Therefore we can see that 
the mass of the lightest excitation is the same as that of a D0-brane 
and identify the one with a D0-brane. Similarly the heavier excitations
can be identified with multiple D0-branes.  

Let us construct the soliton configuration on ${\mb{S}}^2$ in order to
see more clearly that the excitation can be interpreted as a
noncommutative soliton on the fuzzy sphere. On the fuzzy sphere the
spherical harmonics 
$Y_{j,m}$ is represented as the following $p\times p$ matrix
\cite{fuzzy2} 
\ba
(T_{j,m})_{m_1,m_2}&=&(-1)^{\f{p-1}{2}-m_1}\s{2j+1}
\left(
	\begin{array}{ccc}
	\f{p-1}{2} & j & \f{p-1}{2} \\
	-m_1 & m & m_2
	\end{array}
\right),\no
\mbox{Tr}[T_{j,m}T_{j',m'}]&=&\delta_{j,j'}\delta_{m+m',0}(-1)^m,
\ea
where we set $m_1,m_2=-\f{p-1}{2},-\f{p-3}{2},\ddd,\f{p-1}{2}$ and
$(:::)$ denotes the $3j$-symbol. Then one can see that the matrix $T=t^*\mbox{diag}(0^{\f{p-1}{2}-a},1,0^{\f{p-1}{2}+a})$ corresponds to the following tachyon configuration
\ba
T^{(a)}(\theta,\phi)=\f{(-1)^{\f{p-1}{2}-a}}{\s{p}}t^* \sum_{l=0}^{p-1}(2l+1)
\left(
	\begin{array}{ccc}
	\f{p-1}{2} & l & \f{p-1}{2} \\
	-a & 0 & a
	\end{array}
\right)
P_l(\cos\theta), \label{ns}
\ea
where $P_l(\cos\theta)$ denote the Legendre polynomials. The overall
normalization is determined by using that the trace
corresponds to $\f{1}{2\pi\Theta}\int d\phi\sin\theta d\theta$ and the
explicit value of (\ref{the}).
Note that this is independent of $\phi$ and axially symmetric. 
We can construct the more general solution which is not symmetric by performing $U(p)$ rotation.

\begin{figure}[tbp]
\centerline{\epsfbox{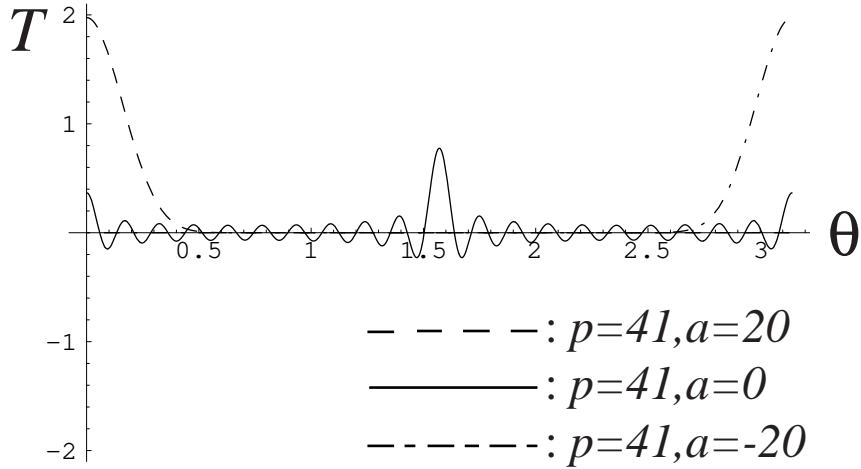}} 
\caption{A plot of the equation (\ref{ns}). Here we set $t^*=1$. The case $(p,a)=(41,20)$ and
$(p,a)=(41,-20)$ correspond to the solitons at the pole of the sphere. 
The case $(p,a)=(41,0)$  corresponds to the one at the equator. }
\label{tachyon}
\end{figure}

When $a=\f{p-1}{2}$ the above can be written as
\ba
T^{(\f{p-1}{2})}(\theta,\phi)=\f{1}{\s{p}}t^* \sum_{l=0}^{p-1}(2l+1)\f{(p-1)!}{\s{(p+l)!(p-l-1)!}}
P_l(\cos\theta).
\ea
From this expression, it is not so difficult to see that in the region $p \gg 1, \theta
\ll 1$ the 
configuration approximates\footnote{We confirmed this up to the
normalization by exact calculations. We checked the overall
normalization in the numerical method.} to the known noncommutative soliton 
\ba
T^{(\f{p-1}{2})}(\theta)\sim 2t^* e^{-\f{p}{2}\theta^2}
\ea
in the
flat space \cite{gop} (see figure \ref{tachyon}).
Similarly we can show
\ba
T^{(-\f{p-1}{2})}(\theta)\sim 2(-1)^{p-1}t^* e^{-\f{p}{2}(\theta-\pi)^2}.
\ea
 From this one can read off the value of the noncommutative parameter for large $p$ as 
\ba
\Theta=\f{2}{p}~. \label{the}
\ea
This value is the same as that obtained in \cite{paw}, where the values of
gauge flux and the $B$-field in the $N\to\infty$ limit were compared with
the analysis given by Seiberg and Witten \cite{sw}. In this
way we have verified that the excitation in this model can be regarded
as a noncommutative soliton on the fuzzy sphere for finite and at least
large $p$. Formally we can also consider the corresponding configuration
(\ref{ns}) for small $p$, but the fuzziness $\delta x\sim\s{\Theta}$ is
comparable to the radius of the sphere and the name ``soliton" does not
seem to be appropriate. 

It is natural to believe that the other cases ($a\neq\pm\f{p-1}{2}$)
are also treated as noncommutative solitons because they have the same
energy. In paticular in the region $p \gg 1, \theta \ll 1, \f{p-1}{2}-a
\ll p$ we 
argue\footnote{We checked this in the numerical method.}
\ba
T^{(a)}(\theta)\sim 2(-1)^{n}t^* L_{n}(p\theta^2)e^{-\f{p}{2}\theta^2},
\ea
where $L_{n}$ is Laguerre polynomial as discussed in \cite{gop} and we defined $n=\f{p-1}{2}-a$.
In other words, the solitons near the pole of sphere can be treated 
approximately as those in the flat space. 

On the other hand,
the solitons with $a\sim 0$
have a peak near the equator $\theta=\f{\pi}{2}$ and is peculiar to the
soliton on the sphere (see figure \ref{tachyon}). Because we cannot
construct a D1-brane in a conformal invariant way, it is natural to
regard such a soliton as a quantum mechanically extended D0-brane. 
As we shall see below that such an object carries more energy than the others
if $N$ is finite. In any case the clear interpretation of such
configurations in conformal field theory is certainly desirable.

In this way we have constructed the noncommutative soliton configurations of the tachyon field on the fuzzy sphere. Each of these has the mass of a D0-brane and obeys the following relation
\ba
\sum_{a=-\f{p-1}{2}}^{\f{p-1}{2}}T^{(a)}(\theta,\phi)=t^*.
\ea
If we consider an excitation $T^{(a_1)}+T^{(a_2)}+\ddd+T^{(a_k)}$ and set $a_i\neq a_j $ for $ i\neq j$, then obviously we get $k$ D0-branes. In the case of $k=p$ we get the original D2-brane. 

We can also show that there are massless fields on the noncommutative soliton
by closely following the discussion of \cite{harvey}.
As we mentioned above the low energy effective action of gauge fields on
the fuzzy sphere was given by \cite{alek3}. 
To include the coupling of tachyon field to gauge fields 
we can replace the derivatives with the covariant one.
In the large $p$ case, the action approaches the flat one
but we should mention that now we deal with the finite dimensional case.
Then it can be shown by following \cite{harvey} that there are two massless
scalars and $(p-1)$ W bosons on 
the lightest noncommutative soliton besides the 
tachyon\footnote{The slightly different approach has been quite recently
proposed by 
\cite{gop2} but the qualitative behavior does not seem to be changed.}.
When the soliton corresponds to $k$ D0-branes, $U(p)$ symmetry breaks into $U(k) \times
U(p-k)$ symmetry. 
It is interesting to investigate more detail but 
it is beyond the scope of this paper. 

Next we discuss the effect of the kinetic term in (\ref{tak}). Such a term is important if $N$ is finite and it breaks $U(p)$ symmetry into $SO(3)$. For $T^{(a)}$ we get
\ba
\mbox{Tr}[\sum_{i=1,2,3}T^{(a)}[X^i,[X^i,T^{(a)}]\ ]\ ]=\f{(t^*)^2}{p^2-1}(2p^2-8a^2-2).
\ea
This means that $a=-\f{p-1}{2}$ and $a=\f{p-1}{2}$ are the lowest
energy configurations for finite $N$. 

 All of the above discussions hold for the D-brane in bosonic string. 
 The exception is  the existence of ``massless"  excitation observed in
 \cite{harvey,das,witten2}, which is peculiar to  non-BPS D-branes. In our case
 this corresponds to $T=\mbox{diag}(0,\ddd,0,2t^*,0,\ddd,0)$. Note that
 the tachyon potential of a non-BPS D-brane is double well and reaches
 its minimum at $T=0,2t^*$ (see figure \ref{doublewell}). Though its
 physical interpretation is not clear also in our case, we can say that
 the interpretation proposed in \cite{das} will not hold in our
 setup. In \cite{das} the authors argued that the tensionless soliton
 corresponds to a BPS D1-brane winding around its core. However in the
 $SU(2)$ WZW model we cannot construct a D1-brane in any way\footnote{
After we submitted the first version of this paper, there
 appeared an argument
\cite{hkl} that the ``massless'' excitation 
is gauge equivalent to the vacuum.}.

As a final task let us consider D2-$\overline{\mbox{D2}}$ systems in the
$N\to\infty$ limit. Here we denote the brane and the antibrane 
by the boundary state $|a\lb$ and $|\bar{b}\lb\ \ (a \geq b)$, respectively. 
As in the previous case we can retain only the double well potential term
and the total energy $E(T)$ is given by\footnote{Unlike the previous
case we do not shift the tachyon field $T$ and the potential have the
symmetry $V(T)=V(-T)$. This is only a matter of convention.} 
\ba
E(T)=\f{1}{G_s}\mbox{Tr}V(T)=M_{D0}\sum_{n=0}^{\infty}c_n \mbox{Tr}\left[(TT^{\dagger})^n+(T^{\dagger}T)^n \right],
\ea
where $c_n\ \ (c_0=1)$ are some constants and $M_{D0}$ denotes the mass
of D0-brane. The tachyon field $T$ is complex $a \times b$ matrix and
$T^{\dagger}$ denotes its hermitian conjugate. The overall normalization
is determined by using the fact that $T=0$ represents the original
brane-antibrane system, namely $E(T=0)=M_{D2-\overline{ D2}}$. Here
$M_{D2-\overline{ D2}}$ denotes the mass of the brane-antibrane system
and is given by 
\ba
M_{D2-\overline{D2}}=(a+b)M_{D0}
 \label{tension}
\ea
in the $N\to\infty$ limit.
Let us define $s^*$ such that $\hat{V}(s)=G_sM_{D0}\sum_{n=0}^{\infty}c_n s^n\
(s\in {\mb{R}}>0)$ take its minimum value at $s^* (>0)$. The point $s=0$ corresponds to the local maximum $\hat{V}(s=0)=G_sM_{D0}$. Then the equation of motion is satisfied if 
\ba
TT^{\dagger}T=s^*T,\ \ \ T^{\dagger}TT^{\dagger}=s^*T^{\dagger}. \label{tt}
\ea 
The same equation is obtained in \cite{witten2,vor}, where the
noncommutative tachyon is discussed in a flat space.
Note that the tachyon field in our case is a finite dimensional matrix.
%This is an finite dimensional analog of the noncommutative tachyon equation 
%in a flat space discussed in
%\cite{witten2,vor} . 

If $a=b$, then the tachyon condensation can be
treated as that on the non-BPS D-brane. 
The maximal
condensation $T=\s{s^*}e^{i\ap}\mbox{diag}(1,\ddd,1)$, where $\ap$ is an 
arbitrary phase, means the decay to the vacuum and we get $\hat{V}(s^*)=0$.
 Then it is easy to see that a tachyon configuration
$T=\s{s^*}e^{i\ap}\mbox{diag}(0^{m},1^{a-m})$ has its energy $E(T)=2m M_{D0}$ and generates $m$ D0-branes and $m$ 
anti D0-branes.

Let us consider the more interesting case $a\neq b$. A series of solutions to (\ref{tt}) are given by
\ba
T^{(l)}=\s{s^*}
\left(
	\begin{array}{cc}
	[1]_{l} &0    \\
	 0 &[0]_{b-l} \\
	 0 & 0 \\
	\end{array}
\right)\ \ \ (l=0,1,\ddd,b),
\ea
where $[a]_{m}$ denotes the $m\times m$ diagonal matrix
$\mbox{diag}(a,a,\ddd,a)$. General solutions are obtained by the further
$U(a)\times U(b)$ gauge rotation. The energy of each solution is given by
\ba
E(T^{(l)})=(a+b-2l)M_{D0}.
\ea
Furthermore the conservation of the RR-charge also guarantees that the total number of D0-branes remains $(a-b)$ in any process. Thus we get 
\be
\begin{array}{ccccc}
 \mbox{The number of D0-branes} &= & \mbox{dim (Ker }T)&= & a-l~,\\
  \mbox{The number of anti D0-branes} &=& \mbox{dim (Ker }T^{\dagger})
   &=& b-l~.
\end{array}
%&& \mbox{The number of D0-branes}=\mbox{dim (Ker }T)=a-l, \no 
%&& \mbox{The number of anti D0-branes}=\mbox{dim (Ker }T^{\dagger})=b-l.
\ee 
If we take the difference between the above equations, we get
\ba
\mbox{Index}(T)=a-b=\int_{{\mb{S}}^2}\f{F^{(a)}-F^{(b)}}{2\pi}, \label{index}
\ea
where $F^{(a)}$ and $F^{(b)}$ denote the gauge fluxes on the world-volume
of the brane and the antibrane, respectively. Their explicit values
are given by (\ref{flux}). 

The relation between the noncommutative tachyon in the brane-antibrane
 system and K-theory of noncommutative algebras has been discussed quite
 recently in \cite{witten2,moore}. Our result offers a special example
 of this, where the (finite dimensional) matrix algebra is involved.
 We argue that the relation (\ref{index}) is regarded as a generalized index
 theorem of noncommutative algebras  (see for example
 \cite{tex1,tex2}). The left-hand side of (\ref{index}) is obtained from
 the vertex operator analysis and the right-hand side
 represents that the D2-brane carries the K-theory charge of D0-brane. 
Therefore we can see that the requirement that this
 relation is consistent with the index theorem ensures the bound state
 interpretation. In order to see the relation
 between the Index($T$) and K-theory charge more clearly,
  it would be interesting to 
investigate the tachyon condensation in the other rational conformal field theories.

\vspace{3ex}

%\begin{center}
\noindent{\Large \bf Acknowledgments}
%\end{center}
\vspace{2ex}\\
We are grateful to Y. Matsuo for useful comments. T.T. is supported by JSPS Research Fellowships for Young Scientists.

%\newpage

\end{document}